\documentclass[12pt]{article}
\usepackage{epsfig}
\usepackage{amsmath}
\usepackage{hhline}
\usepackage{amssymb}
\usepackage{times}
\usepackage{cite}

\newlength{\dinwidth}
\newlength{\dinmargin}
\begin{document}  
\newcommand{\pom}{{I\!\!P}}
\newcommand{\reg}{{I\!\!R}}
\newcommand{\slowpi}{\pi_{\mathit{slow}}}
\newcommand{\fiidiii}{F_2^{D(3)}}
\newcommand{\fiidiiiarg}{\fiidiii\,(\beta,\,Q^2,\,x)}
\newcommand{\n}{1.19\pm 0.06 (stat.) \pm0.07 (syst.)}
\newcommand{\nz}{1.30\pm 0.08 (stat.)^{+0.08}_{-0.14} (syst.)}
\newcommand{\fiidiiiful}{F_2^{D(4)}\,(\beta,\,Q^2,\,x,\,t)}
\newcommand{\fiipom}{\tilde F_2^D}
\newcommand{\ALPHA}{1.10\pm0.03 (stat.) \pm0.04 (syst.)}
\newcommand{\ALPHAZ}{1.15\pm0.04 (stat.)^{+0.04}_{-0.07} (syst.)}
\newcommand{\fiipomarg}{\fiipom\,(\beta,\,Q^2)}
\newcommand{\pomflux}{f_{\pom / p}}
\newcommand{\nxpom}{1.19\pm 0.06 (stat.) \pm0.07 (syst.)}
\newcommand {\gapprox}
   {\raisebox{-0.7ex}{$\stackrel {\textstyle>}{\sim}$}}
\newcommand {\lapprox}
   {\raisebox{-0.7ex}{$\stackrel {\textstyle<}{\sim}$}}
\def\gsim{\,\lower.25ex\hbox{$\scriptstyle\sim$}\kern-1.30ex%
\raise 0.55ex\hbox{$\scriptstyle >$}\,}
\def\lsim{\,\lower.25ex\hbox{$\scriptstyle\sim$}\kern-1.30ex%
\raise 0.55ex\hbox{$\scriptstyle <$}\,}
\newcommand{\pomfluxarg}{f_{\pom / p}\,(x_\pom)}
\newcommand{\dsf}{\mbox{$F_2^{D(3)}$}}
\newcommand{\dsfva}{\mbox{$F_2^{D(3)}(\beta,Q^2,x_{I\!\!P})$}}
\newcommand{\dsfvb}{\mbox{$F_2^{D(3)}(\beta,Q^2,x)$}}
\newcommand{\dsfpom}{$F_2^{I\!\!P}$}
\newcommand{\gap}{\stackrel{>}{\sim}}
\newcommand{\lap}{\stackrel{<}{\sim}}
\newcommand{\fem}{$F_2^{em}$}
\newcommand{\tsnmp}{$\tilde{\sigma}_{NC}(e^{\mp})$}
\newcommand{\tsnm}{$\tilde{\sigma}_{NC}(e^-)$}
\newcommand{\tsnp}{$\tilde{\sigma}_{NC}(e^+)$}
\newcommand{\st}{$\star$}
\newcommand{\sst}{$\star \star$}
\newcommand{\ssst}{$\star \star \star$}
\newcommand{\sssst}{$\star \star \star \star$}
\newcommand{\tw}{\theta_W}
\newcommand{\sw}{\sin{\theta_W}}
\newcommand{\cw}{\cos{\theta_W}}
\newcommand{\sww}{\sin^2{\theta_W}}
\newcommand{\cww}{\cos^2{\theta_W}}
\newcommand{\trm}{m_{\perp}}
\newcommand{\trp}{p_{\perp}}
\newcommand{\trmm}{m_{\perp}^2}
\newcommand{\trpp}{p_{\perp}^2}
\newcommand{\alp}{\alpha_s}

\newcommand{\alps}{\alpha_s}
\newcommand{\sqrts}{$\sqrt{s}$}
\newcommand{\LO}{\mathcal{O}(\alpha_s^0)}
\newcommand{\Oa}{\mathcal{O}(\alpha_s)}
\newcommand{\Oaa}{O(\alpha_s^2)}
\newcommand{\PT}{p_{\perp}}
\newcommand{\JPSI}{J/\psi}
\newcommand{\sh}{\hat{s}}
\newcommand{\uh}{\hat{u}}
\newcommand{\MP}{m_{J/\psi}}
\newcommand{\PO}{I\!\!P}
\newcommand{\xbj}{x}
\newcommand{\xpom}{x_{\PO}}
\newcommand{\my}{M_{_Y}}
\newcommand{\mx}{M_{_X}}
\newcommand{\dstar}{D^{*\pm}}
\newcommand{\zpom}{z_{\PO}}
\newcommand{\zpomobs}{z_{\PO}^{obs}}
\newcommand{\ptgp}{p^*_{T, D^*}}
\newcommand{\ptlab}{p_{T, D^*}}
\newcommand{\etalab}{\eta_{D^*}}
\newcommand{\ttbs}{\char'134}
\newcommand{\xpomlo}{3\times10^{-4}}  
\newcommand{\xpomup}{0.05}  
\newcommand{\dgr}{^\circ}
\newcommand{\pbarnt}{\,\mbox{{\rm pb$^{-1}$}}}
\newcommand{\gev}{\,\mbox{GeV}}
\newcommand{\WBoson}{\mbox{$W$}}
\newcommand{\fbarn}{\,\mbox{{\rm fb}}}
\newcommand{\fbarnt}{\,\mbox{{\rm fb$^{-1}$}}}
%
%
\newcommand{\qsq}{\ensuremath{Q^2} }
\newcommand{\gevsq}{\ensuremath{\mathrm{GeV}^2} }
\newcommand{\et}{\ensuremath{E_t^*} }
\newcommand{\rap}{\ensuremath{\eta^*} }
\newcommand{\gp}{\ensuremath{\gamma^*}p }
\newcommand{\dsiget}{\ensuremath{{\rm d}\sigma_{ep}/{\rm d}E_t^*} }
\newcommand{\dsigrap}{\ensuremath{{\rm d}\sigma_{ep}/{\rm d}\eta^*} }
\def\Journal#1#2#3#4{{#1} {\bf #2} (#3) #4}
\def\NCA{\em Nuovo Cimento}
\def\NIM{\em Nucl. Instrum. Methods}
\def\NIMA{{\em Nucl. Instrum. Methods} {\bf A}}
\def\NPB{{\em Nucl. Phys.}   {\bf B}}
\def\PLB{{\em Phys. Lett.}   {\bf B}}
\def\PRL{\em Phys. Rev. Lett.}
\def\PRD{{\em Phys. Rev.}    {\bf D}}
\def\ZPC{{\em Z. Phys.}      {\bf C}}
\def\EJC{{\em Eur. Phys. J.} {\bf C}}
\def\CPC{\em Comp. Phys. Commun.}

\begin{titlepage}

\begin{flushleft}
DESY 01-105 \hfill ISSN 0418-9833 \\
July 2001
\end{flushleft}

\vspace{2cm}

\begin{center}
\begin{Large}

{\bf  {\boldmath $\dstar$} Meson Production in
Deep--Inelastic Diffractive Interactions at HERA }

\vspace{2cm}

H1 Collaboration

\end{Large}
\end{center}

\vspace{2cm}

\begin{abstract}
\noindent
A measurement is presented of the cross section for
$\dstar$ meson production in diffractive deep--inelastic 
scattering for the first time at HERA. 
The cross section is given for the
process $ep \rightarrow eXY$, where the system $X$
contains at least one $\dstar$ meson and is separated 
by a large rapidity gap from a low mass proton remnant system $Y$.
The cross section is presented in the diffractive deep--inelastic region
defined by $2< Q^2 < 100 \ {\rm GeV^2}$, $0.05 < y < 0.7$, 
$\xpom < 0.04$, $M_Y < 1.6  \ {\rm GeV}$ and $|t| < 1 \ {\rm GeV^2}$. 
The $\dstar$ mesons are restricted to the range $\ptlab > 2 \ {\rm GeV}$ and 
$|\etalab | < 1.5$.
The cross section is found to be $246 \pm 54 \pm 56 \ {\rm pb}$ and forms 
about $6\%$ of the corresponding inclusive $\dstar$ cross section. 
The cross section is presented as a function of 
various kinematic variables, including $\zpomobs$ which 
is an estimate of the
fraction of the momentum of the diffractive exchange carried by the 
parton entering the hard-subprocess.
The data show a large component of the cross section at low
$\zpomobs$ where the contribution
of the Boson--Gluon--Fusion process is expected to dominate.
The data are compared with several QCD--based calculations.
\end{abstract}

\vspace{1.5cm}

\begin{center}
Submitted to {\em Phys. Lett. B.}
\end{center}

\end{titlepage}

%
%

\begin{flushleft}

C.~Adloff$^{33}$,              
V.~Andreev$^{24}$,             
B.~Andrieu$^{27}$,             
T.~Anthonis$^{4}$,             
V.~Arkadov$^{35}$,             
A.~Astvatsatourov$^{35}$,      
A.~Babaev$^{23}$,              
J.~B\"ahr$^{35}$,              
P.~Baranov$^{24}$,             
E.~Barrelet$^{28}$
% ...

\bigskip{\it
% ...
$ ^{4}$ Inter-University Institute for High Energies ULB-VUB, Brussels;
          Universitaire Instelling Antwerpen, Wilrijk; Belgium$^{ c}$ \\
% ...
$ ^{23}$ Institute for Theoretical and Experimental Physics,
          Moscow, Russia$^{ l}$ \\
 $ ^{24}$ Lebedev Physical Institute, Moscow, Russia$^{ e,h}$ \\
% ...
$ ^{27}$ LPNHE, Ecole Polytechnique, IN2P3-CNRS, Palaiseau, France \\
 $ ^{28}$ LPNHE, Universit\'{e}s Paris VI and VII, IN2P3-CNRS,
          Paris, France \\
% ...
$ ^{33}$ Fachbereich Physik, Bergische Universit\"at Gesamthochschule
          Wuppertal, Wuppertal, Germany \\
% ...
$ ^{35}$ DESY, Zeuthen, Germany \\
% ...
$ ^c$ Supported by FNRS-NFWO, IISN-IIKW \\
% ...
$ ^e$ Supported by the Deutsche Forschungsgemeinschaft \\
% ...
$ ^h$ Supported by Russian Foundation for Basic Research
      grant no. 96-02-00019 \\
% ...
$ ^l$ Partially Supported by Russian Foundation
      for Basic Research, grant    no. 00-15-96584 \\
}

\end{flushleft}

\newpage
\section{Introduction}

The observation of events with a large rapidity gap in the distribution of the 
final state hadrons at HERA~\cite{first} allows the nature of colour singlet exchange
in strong interactions to be investigated.  Colour singlet
exchange interactions have been successfully modelled~\cite{collins} in terms of
phenomenological Regge theory~\cite{regge} and, at high energy, 
are attributed to diffractive or
pomeron exchange.  HERA allows the partonic nature of diffraction
to be investigated in deep--inelastic scattering (DIS) using the
virtual photon as a probe.

The inclusive diffractive DIS structure function $F_2^{D}$
is directly
sensitive to the quark content of the diffractive exchange~\cite{H1:f2d394,f2dzeus}.  
Information about the gluon content 
can be inferred indirectly from scaling violations. 
However, the measurement of the hadronic final state in diffraction gives further, more 
direct, information about the gluon content~\cite{H1:diffhfs,diffhfs-zeus,
djzeus+H1:d2j94,H1:d2j00}.
The production of open charm is expected to be particularly sensitive to the
gluon content because studies in inclusive DIS reveal that the dominant
contribution comes from the boson--gluon--fusion (BGF)
mechanism~\cite{f2c_bgf}.  
The presence of the hard scale, provided by the 
charm quark mass, allows a variety of perturbative QCD-based models 
of diffraction to be tested.

This article describes the measurement of diffractive open charm production in DIS
at HERA, which was performed using the H1 detector. Measurements of the total $\dstar$ cross 
section and of differential distributions which explore the dynamics of 
diffractive charm production are presented. The ratio of the
diffractive $\dstar$ cross section 
to the inclusive $\dstar$ cross section is also measured.

The paper is organised as follows. The kinematics of diffractive DIS are
introduced 
in section \ref{section:kinematics}.
The different theoretical approaches to diffractive charm production 
are summarised in section \ref{section:models}.
In section \ref{section:experiment}, the H1 detector, the data selection, 
the cross section measurement procedure and the
evaluation of the systematic uncertainties are explained.  The results, in the form of
the total and differential cross sections are presented and 
discussed in section \ref{section:results}.
 
\section{Kinematics}
\label{section:kinematics}

\begin{figure}[t]
\centering \epsfig{file=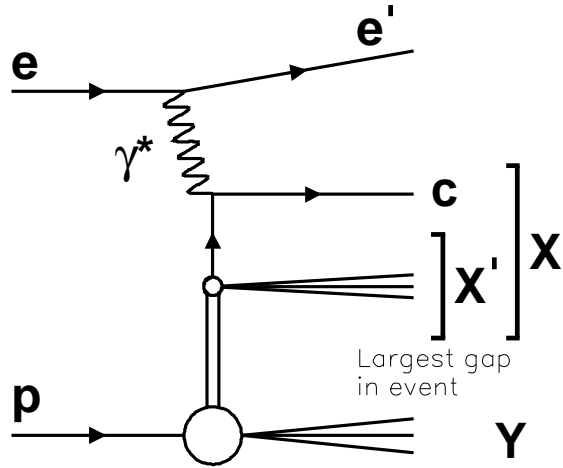,width=0.5\linewidth}
\caption{ \em The process under study in this article is $ep \rightarrow
eXY \rightarrow e(\dstar X')Y$.
The electron ($e$) couples to a photon ($\gamma^{\star}$) which interacts with
the proton ($p$) via colour singlet exchange, producing two distinct
final state hadronic systems $X$ and $Y$.  The systems $X$ and $Y$ are
separated by the largest gap in rapidity in the final state hadrons.}
\label{kinefig}
\end{figure}

The process studied in this paper is $ep \rightarrow eXY \rightarrow e(\dstar X')Y$ and is 
shown in figure \ref{kinefig}. 
The electron produces
a virtual photon $\gamma^{\star}$ (with four--momentum $q$) which interacts with the 
proton (with four--momentum $P$).   
If the interaction takes place via colour singlet exchange,  
the photon and proton dissociate into two distinct hadronic systems $X$ and $Y$, with
invariant masses $\mx$ and $\my$, respectively. The system $Y$ 
is that which is closest to the outgoing proton direction. 
In the case where $\mx$ and $\my$ are small compared with the photon-proton centre 
of mass energy $W$, the two systems are separated by a large rapidity gap.
In addition to the standard DIS 
kinematic variables $Q^2$, $y$ and Bjorken $x$ the following variables are defined
\begin{equation}
  \label{xpomtbeta}
 \xpom = \frac{q \cdot (P - p_Y)}{q \cdot P} \ ; \qquad t = (P-p_Y)^2 \ ; \qquad
\beta = \frac{Q^2}{2q\cdot (P-p_Y)} = \frac{x}{\xpom},
\end{equation}
where $p_Y$ is the four--momentum of $Y$.
The quantity $\xpom$ may be interpreted as the longitudinal momentum fraction,
with respect to the incoming proton, of the colourless exchange and $t$ is
the squared four--momentum transferred at the proton vertex. In the analysis
presented in this paper, $t$ and $\my$ are constrained to be small by the
experimental selection and are integrated over implicitly.

In a QCD interpretation in which a partonic structure is ascribed to the 
colourless exchange
the lowest order 
(i.e. $\LO$ ) contribution to the diffractive cross section 
in the proton infinite momentum frame is quark 
scattering ($\gamma^{\star} q \rightarrow q$). In this case $\beta$
can be interpreted as the fractional longitudinal momentum of the
exchange carried by the struck quark.    
The $\Oa$ contributions are the BGF  
($\gamma^{\star} g \rightarrow q\bar{q}$) 
and 
QCD-Compton ($\gamma^{\star} q \rightarrow qg$) processes. 
In the $\Oa$ case the invariant mass squared $\hat{s}$ of the partons emerging from the 
hard subprocess is non-zero. Therefore, the quantity $\zpom$ is introduced
\begin{equation}
  \label{zpom}
 \zpom = \beta \cdot \left(1 + \frac{\hat{s}} {Q^2}\right) 
\end{equation}
which corresponds to the longitudinal momentum fraction of the colourless exchange
carried by the parton (quark or gluon) 
which enters the hard interaction.

\section{Models of Diffractive {\boldmath $\dstar$} Production}
\label{section:models}

A detailed description of the models used in this paper is given in 
\cite{H1:d2j00}. A brief summary focusing on the production of open
charm in each of the models is given here.  For ease of
comparison with the data in this paper 
the models are divided into three groups: 
the `resolved pomeron' model, `2-gluon exchange' models and 
`soft colour neutralisation' models. 

In the `resolved pomeron' model \cite{respom}  the diffractive cross section 
factorises into deep--inelastic scattering from the pomeron 
and a pomeron flux factor, motivated by Regge theory, which describes 
the probability for finding a pomeron in the proton. 
Within this picture, the partonic content of the pomeron has been determined 
by QCD analyses of HERA diffractive data \cite{H1:f2d394,actw}.
The parton distributions, obtained from
fits to the data, contain a dominant gluon distribution.  
Open charm is produced in the resolved pomeron model
by the BGF process, where the photon interacts with a gluon of the pomeron
carrying a fraction $\zpom$ of the pomeron longitudinal momentum. 

In `2-gluon exchange' models diffractive DIS is studied 
in the proton rest frame 
by considering $q\overline{q}$ and $q\overline{q}g$ 
photon fluctuations as colour dipoles
scattering off the proton.  
Open charm can be produced when the 
photon fluctuates into $c\overline{c}$ or $c\overline{c}g$ states.
The simplest realisation of net colour singlet exchange 
between these partonic fluctuations and the proton at the parton level 
is a pair of 
gluons with opposite colour \cite{lownussinov}.  
In perturbative QCD, the 
cross section for 2-gluon
exchange is related to the square of the $k_{T}$--unintegrated gluon density of the 
proton $\mathcal{F}(x,k_T^2)$ \cite{bfkl,off_shell_me},
where $k_T$ is the parton transverse momentum relative to the proton direction.
In the `saturation' model \cite{sat} the calculation of the
$q\overline{q}g$ cross section is
made under the assumption of strong $k_T$ ordering of the final state partons,
where $k^{(g)}_T \ll k^{(q,\overline{q})}_T$.  In an alternative
approach \cite{bartelsqq,bartelsqqg} (hereafter referred to as `BJLW')  
the calculation of the $q\overline{q}g$ final 
state includes configurations without strong $k_T$ ordering.  
In this model all
outgoing partons are required to have high $k_T$ and the minimum value for 
the final state gluon transverse momentum $k_{T,g}^{cut}$ is a free parameter which
can be tuned to describe the data.
The 2-gluon exchange calculations are performed under the
assumption of low $\xpom$ ( $\xpom < 0.01$ ) to avoid contributions 
from secondary reggeon exchanges which correspond to quark exchange 
in the models. 

In `soft colour neutralisation' models an alternative approach to diffractive DIS 
is given which leads to very similar properties of inclusive and diffractive
DIS final states.  In the soft colour interaction model (SCI) \cite{sci}, 
open charm is produced via BGF from the gluon distribution of the proton. 
It is then assumed that the partons produced in the hard interactions can
exchange soft gluons with the background colour field of the incoming 
proton leaving all momenta unchanged. Large rapidity gap events may be
produced this way when the soft interactions lead to a net colour singlet
exchange. The probability for soft colour exchange is assumed to be
independent of the kinematics of the hard scattering process.

The generalised area law (GAL)\cite{gal} approach is a modification of the 
Lund String model \cite{lund}. The production mechanism for open charm 
is similar to that in the SCI model except that
it is formulated in terms of interactions between the colour strings
connecting the partons in an event. In this model the probability for a
soft colour interaction is not constant but is exponentially suppressed
by the difference between the areas in momentum space spanned by the 
strings before and after the colour rearrangement.

The `semi-classical' model \cite{semicl} is a non-perturbative model 
based on the dipole approach.  In the proton rest frame the photon 
fluctuations scatter
off a superposition of soft colour fields representing the proton. In this
approach, the $q\overline{q}g$ fluctuation is expected to be
dominant for open 
charm production\cite{semiclcharm}. If the gluon is 
the lowest $k_T$ parton, then the contribution can be related to BGF in 
the proton infinite momentum frame.

Comparison of the data with the `resolved pomeron', the `2-gluon
exchange' 
and the `semi-classical' models 
is facilitated by their implementation within the RAPGAP Monte Carlo generator 
\cite{rapgap}.
The predictions of the SCI and GAL models are calculated using the AROMA Monte
Carlo generator \cite{aromsci}. 
The cross section predictions in this article are all calculated assuming a charm quark 
mass $m_c=1.5 \ {\rm GeV}$. For the hadronisation fraction $f(c \rightarrow \dstar)$ the 
value $0.233 \pm 0.010 \pm 0.011$ \cite{hadfrac} is used. The momentum fraction of the
charm quark carried by the $\dstar$ is calculated using the Peterson 
{\em et al.} model \cite{peterson} 
with the fragmentation parameter $\epsilon~=~0.078$.

\section{Experimental Procedure}
\label{section:experiment}

The data presented in this analysis were collected 
over the years 1996 and 1997, when HERA collided positrons with energy
$E_e = 27.5 \ {\rm GeV}$ with protons of energy $E_p = 820 \ {\rm GeV}$.  
Requiring all essential detector components to be operational the available 
integrated luminosity is $19.1 {\rm pb}^{-1}$.   Further details of this analysis 
beyond those given here can be found in \cite{H1:Stefan}.

\subsection{The H1 Detector}

A short overview of the detector components most relevant for the present analysis
is given here. A detailed description of the H1 detector can be found in \cite{H1:det}. 
The $z$--axis of the H1 detector is taken along the beam direction such that 
positive $z$ values refer to the direction of  the outgoing proton beam, 
referred to as the `forward' direction.

Charged particles emerging from the interaction region are measured by the central tracking
device (CTD) in the range $-1.5 < \eta < 1.5$\footnote{The pseudorapidity 
$\eta$ of an object detected with polar angle $\theta$ is defined as
$\eta = - \ln \ \tan (\theta / 2)$.}.  The CTD comprises two large
cylindrical central jet drift chambers (CJC) 
and two $z$ chambers situated concentrically around the beam-line 
within a solenoidal magnetic field of 1.15 T. The resolution achieved by the 
CTD is $\sigma(p_T)/p_T \simeq 0.01 p_T/\mathrm{GeV}$. 
The CTD also provides triggering
information  based on track segments in the $r-\phi$ plane 
from the CJC and the position of the vertex using a double 
layer of multi-wire proportional chambers (MWPC).
The energies of final state particles are measured in the Liquid Argon (LAr) calorimeter
which surrounds the tracking chambers and covers the range $-1.5 < \eta < 3.4$. 
The backward region ($-4.0 < \eta < -1.4$) is covered by
a lead--scintillating fibre calorimeter (SPACAL~\cite{spacal}) with electromagnetic and
hadronic sections.  
In front of the SPACAL, the Backward Drift Chamber
(BDC) \cite{bdc} provides track segments of charged particles. 

Detectors close to the beam pipe in the direction of the outgoing proton are used in 
the selection of large rapidity gap events.  These are the 
Forward Muon Detector (FMD), the Proton Remnant
Tagger (PRT) and the Plug calorimeter (PLUG).
The FMD is located at $z=6.5 \ 
\mathrm{m}$ and covers the pseudorapidity range $1.9<\eta<3.7$
directly. The PLUG allows energy measurements to be made over the
range $3.5<\eta <5.5$.
Particles produced at larger $\eta$ can also be detected
because of secondary scattering with the beam-pipe. The PRT, a set of
scintillators surrounding the beam pipe at $z=26 \ \mathrm{m}$, can
tag hadrons in the region $6.0 \ \lapprox \ \eta \ \lapprox \ 7.5$.

\subsection{Event Selection}

The events were triggered by an electromagnetic energy cluster in the SPACAL, 
in coincidence with a charged track signal 
from both the MWPC and the CJC.  
The positrons are identified in the SPACAL as clusters with energy 
$E'_{e} > 9 $~{\rm GeV} which
have properties consistent with electromagnetic deposition, 
and for which the centre of gravity of the 
cluster matches a charged track segment in the BDC to within
$2.5 \ {\rm cm}$.
The selected events are also required to have a reconstructed vertex from
the CTD within 
$\pm 35\ {\rm cm}$ of the nominal vertex. In order to suppress events with 
initial
state photon radiation the summed $E-p_z$ of the event
calculated using all reconstructed final state particles,
including the positron, is required to be greater than $35 \ {\rm GeV}$.
The kinematic region covered by the measurement
is $2<Q^2<100 ~{\rm GeV^2}$ and 
$0.05<y<0.7$. To minimise the correction due to QED radiative effects, 
$Q^2$, $y$ and $x$ are reconstructed 
from the energy and angle $\theta'$ of the scattered positron and the hadronic final state
using the `$\Sigma$ method' \cite{sigmameth}.

Diffractive events are 
selected experimentally by the absence of activity in the outgoing
proton region.
No signal above noise thresholds is allowed in the FMD, the PRT, the PLUG
and the most forward part ($\eta > 3.3$) of the LAr calorimeter.
This ensures that there
is a large rapidity gap covering at least $3.3 < \eta \le 7.5$ between the 
photon dissociation system $X$ and the proton remnant system $Y$.
Monte Carlo studies show that the absence of particles in the 
detectors close to the beam pipe restricts 
the mass of the proton remnant system to $\my < 1.6 \ \rm{GeV}$ 
and the momentum transfer to the proton to $|t| < 1 \ \rm {GeV}^2$. 

The four--momentum of the system $X$, which is well contained in the central detector, 
is reconstructed using 
information from the LAr and SPACAL calorimeters together with the CJC \cite{fscomb}.
The variable
$\xpom$ is calculated from
\begin{equation}
\label{xpomandbeta}
        \xpom = \frac {\sum_{ X + e'}(E + p_z)} {2 E_p} \ 
\end{equation}

where $E$ and $p_z$ are the energy and longitudinal momentum of each final state particle
in the laboratory frame, and the sum runs over the scattered positron $e'$ and all
detected particles in the photon dissociation system $X$. The quantity $\beta$
is calculated from $\beta = x / \xpom$.
The cross section is restricted to the range $\xpom<0.04$ to suppress 
contributions from non-diffractive scattering and secondary reggeon
exchanges. 
  
In this paper an hadronic observable $\zpomobs$ is constructed which is 
analogous to $x_{g}^{obs}$ for inclusive $\dstar$ production which was
measured in \cite{xglu}.
In the resolved pomeron picture $\zpomobs$ is an
approximation to the momentum fraction $\zpom$ of the pomeron 
carried by the interacting gluon (see equation \ref{zpom}). 
The observable $\zpomobs$ is defined as
\begin{equation}
\label{zpomdstar}
 \zpomobs =  \frac{ M_{c\bar{c}}^2 +Q^2}{\mx^2+Q^2}.
\end{equation}
where $M_{c\bar{c}}^2$ is a hadron level estimate of $\hat{s}$
which is constructed
from the scattered positron and the $D^{*\pm}$ meson
in an identical manner to that used for the gluon momentum fraction $x_{g}^{obs}$
in \cite{xglu}. 
Monte Carlo simulations show that the resolution in the
hadronic variable $\zpomobs$ is
approximately $30\%$, and that there is a good correlation between
$\zpomobs$ and $\zpom$ as calculated from the kinematics of the outgoing
partons. 
The variable $\zpomobs$ can be interpreted as the
fraction of the energy of the system $X$ which is carried by 
the $c\bar{c}$ pair emerging from the hard scattering.

\subsection{Reconstruction of {\boldmath $D^{*\pm}$} Mesons}
\label{trackbit}

The $\dstar$ mesons are reconstructed using the $\dstar-D^0$ mass difference method \cite{dmmethod}
in the decay channel
\begin{equation} 
\dstar \rightarrow D^0 \ \pi^{+}_{slow} \rightarrow ( K^{-} \ \pi^{+}) \ \pi^{+}_{slow} 
\ (+ c.c.),
\label{reaction}
\end{equation}
which has a branching fraction of 
$2.59\%$~\cite{brref}.  
The reconstruction method is detailed in \cite{H1:f2c01}.
The decay products are detected in the CTD
and are required to have a transverse momentum $p_T$
of at least $140 \ {\rm MeV}$ for the
$\pi_{slow}$ and $250 \ {\rm MeV}$ for both the $K$ and $\pi$.

\begin{figure}[h]
\begin{center}
\mbox{\epsfig{figure=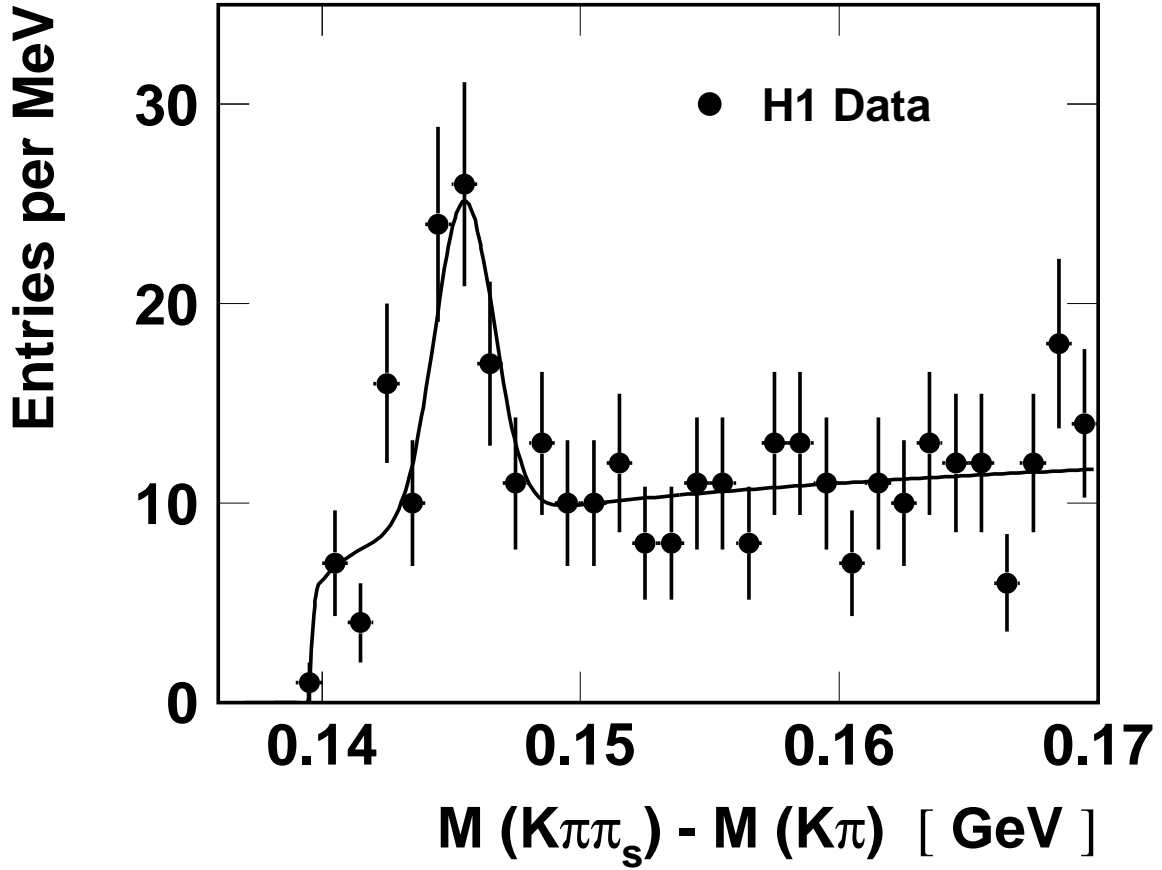,angle=0,width=1.0\textwidth}}
\end{center}
\caption{ \em  Distribution of the mass difference 
$\Delta M=M(K^{\mp} \pi^{\pm} \pi_{slow}^{\pm})-M(K^{\mp}\pi^{\pm})$, with a curve fitted to the 
form $a(\Delta M-M_{\pi})^{b} +$ Gaussian.}
\label{deltam}
\end{figure}

The invariant mass of the $K\pi$ combination has to be consistent with the 
$D^0$ mass within $\pm 80 \ {\rm  MeV}$.  
After cuts on the direction ($|\eta(K \pi \pi)| < 1.5$) and transverse momentum 
($p_T(K \pi \pi) > 2 \ {\rm GeV}$), the mass difference distribution
$\Delta M=M(K \pi \pi_{slow})-M(K \pi)$ is plotted in 
figure~\ref{deltam}. 
The number of $\dstar$ candidates is determined by fitting the histogram 
in figure~\ref{deltam} with a Gaussian distribution for the signal plus a 
background function $a(\Delta M-M_{\pi})^b$, where $M_{\pi}$ denotes the mass of 
the pion.
The position and width of the Gaussian are fixed to values taken from a higher 
statistics sample of events where no diffractive cuts were
applied~\cite{H1:f2c01}.  
The normalisation of the Gaussian and the background parameters 
$a$ and $b$ are allowed to vary. 
The resulting number of detected $\dstar$ mesons is $46 \pm 10$. 

\subsection{Cross Section Measurement} 

Monte Carlo simulations are used to correct the data for the effects of losses 
and migrations due to the finite resolution of the H1 detector.  The 
efficiency is calculated by running the H1 detector 
simulation program on a sample of $\dstar$ events from the diffractive Monte 
Carlo generator RAPGAP \cite{rapgap} in the resolved 
pomeron mode with the $t$ dependence of the cross section parameterised as
$e^{-6|t|}$. The RAPGAP program is used to model events  which contain an
elastic proton ($\my=m_p$) in the kinematic range $\xpom < 0.1$. 
Migrations from $\xpom > 0.1$ or from large values of $M_Y$ ($M_Y>5 \ \rm\ GeV$)
are modelled by using a simulation of  
the heavy quark generator AROMA~\cite{aromsci} in the inclusive mode.
The contribution is of the order $5\%$ to the selected sample of events. 
An additional correction of $-8\% \pm 6\%$ is applied to account
for the net smearing across the $\my=1.6 \ \rm GeV$ boundary.
Since only elastically scattered protons have been simulated in
RAPGAP, this correction is evaluated using the proton dissociation 
simulation in the DIFFVM \cite{diffvm} generator\footnote{For the 
correction, it is assumed that the
ratio of diffractive proton elastic to diffractive proton 
dissociative interactions is $1:1$.}.
A further correction of $+4\%\pm1\%$ takes into account
diffractive events rejected due to fluctuations in the noise level in the
FMD. This correction is estimated directly from the data, using a sample of 
randomly selected events, not correlated with a physics trigger.
An additional source of background is the contribution of
reflections in the $D^0$ mass window, coming from $D^0$ channels 
other than that defined in equation \ref{reaction}.
They are estimated, from simulations using the AROMA Monte Carlo,
to be $3.5\%$ \cite{H1:Stefan}.
The contribution from photoproduction background is found to be 
negligible.  
QED radiative corrections were calculated to be approximately $2\%$ using the
RAPGAP program interfaced to HERACLES \cite{heracles}.

\subsection{Systematic Uncertainties}

The following sources of systematic error are taken into account

\begin{itemize}
\item
The uncertainty in the physics model for $\dstar$ production used
to compute the efficiency corrections is estimated 
by varying the shapes of the kinematic distributions in the
simulations beyond the limits imposed by previous measurements or the 
present data. This is done by reweighting the $\xpom$ distribution
to that observed in data; the $\beta$ distribution by 
$(1 \pm 0.3\beta)$ and the $t$ distribution by $e^{\pm 2t}$.
The resulting systematic uncertainties on the cross section
measurements range between 
$10 \%$ and $20\%$ with the
largest contribution originating from the variation of the $\xpom$ distribution.
The uncertainties are verified using simulations of models with different 
underlying kinematic distributions.

\item
The total uncertainty due to the 
reconstruction efficiency, mass and momentum resolution  of the 
central tracker for the three tracks was estimated in the analysis of the
inclusive DIS $\dstar$ cross section to be  $^{+9\%}_{-4\%}$~\cite{H1:f2c01}. 

\item
An error of $8\%$ is found by varying the
details of the fitting procedure used to obtain the number
of $D^{*\pm}$ mesons.

\item
The uncertainty in the correction due to the smearing of events across the 
boundary $M_Y=1.6 \ \rm\ GeV$ is estimated by varying in the DIFFVM
simulation:  
the efficiency of the forward detectors, the assumed $\my$ distribution,
the ratio of double to single dissociation between 0.5 and 2 and  
the assumed $t$ dependence for double dissociation. 
This contributes $ 6\%$ to the systematic error.   

\item
The uncertainty in the trigger efficiency gives a contribution of $5\%$ 
to the systematic error.

\item
The uncertainty due to the assumed charm fragmentation scheme 
is estimated by using 
parameterisations of the Peterson model 
and the standard Lund string model in JETSET \cite{lund}.
This leads to an average uncertainty of $5\%$ in the cross sections.

\item
The uncertainty in the correction due to QED radiative effects is estimated as $ 3\%$.

\item
The number of events migrating
into the sample from $\xpom>0.1$ or $M_Y>5 \ \rm GeV$ is varied by $\pm 50\%$,
leading to an average systematic error of $3\%$.

Other sources of systematic error are the uncertainty in the 
measured energy and angle of the 
scattered positron, uncertainties in the hadronic 
energy scale of the liquid argon and SPACAL calorimeters, the uncertainty in the 
luminosity measurement, the uncertainty on the fraction of events lost
due to noise in the FMD and the uncertainty in 
the branching ratio for the measured decay channel.
Each of them is responsible for an error of no more than $2.5\%$.
\end{itemize}

The total systematic error for each point has been obtained by adding
all individual contributions in quadrature. It ranges between $20\%$
and $30\%$ and for most data points is similar in magnitude to the
statistical error.

\section{Results}
\label{section:results}

The total $\dstar$ production cross section for the kinematic region 
$2 < Q^2 < 100 ~{\rm GeV^2}$, 
$0.05 < y < 0.7$, 
$\xpom < 0.04$, 
$M_Y < 1.6 ~{\rm GeV}$, 
$|t| < 1 ~{\rm GeV^2}$, 
$\ptlab > 2 \ {\rm GeV}$ and 
$|\etalab|< 1.5$ is
\begin{equation}
 \sigma(ep \rightarrow e(\dstar X')Y)=
 246 \pm 54 \ {\rm (stat.)} \pm 56 \ {\rm (syst.)} \ {\rm pb} \ .
\end{equation}

The ratio of the diffractive $\dstar$ cross section
to the inclusive $\dstar$ cross section measured in the same kinematic range 
defined in terms of $Q^2$, $y$, $\ptlab$ and $\etalab$ is found to be 
\begin{eqnarray}
  5.9 \pm 1.1 \ {\rm (stat.)} \pm 1.1 \ {\rm (syst.)} \%,
\end{eqnarray}
where the inclusive $\dstar$ cross section has been determined
as in~\cite{H1:f2c01}.
The error in the ratio is dominated by the uncertainties pertaining to the
measurement of the diffractive cross section.

\begin{table}[h]
\begin{center}
\begin{tabular}{|c|c||c|} \hline
\multicolumn{2}{|c||} {Model}   & cross section (${\rm pb}$) \\ \hline
resolved&H1 fit 2   & $368$ \\ 
pomeron &H1 fit 3   & $433$ \\ 
        &ACTW fit D & $481$ \\ \hline 
soft colour & SCI  & $203$ \\ 
neutralisation & GAL & $328$ \\ 
& semi-classical   & $196$ \\  \hline 
 \multicolumn{2}{|c||} {H1 Data} & 
{$246 \pm 54 \ {\rm (stat.)} \pm 56 \ {\rm (syst.)} \ {\rm pb}$} \\ \hline
\end{tabular}
\end{center}
\caption{\em The predictions for the total diffractive $\dstar$ 
cross section for two groups of models: the resolved pomeron
and soft colour neutralisation approaches. The bottom row shows the cross
section measured in the data.}
\label{table:model}
\end{table}

In table \ref{table:model} the total cross section is compared with
some of the 
phenomenological models discussed in section \ref{section:models}. 
The first three rows of table \ref{table:model} show the predictions for
the cross section for three different sets of parton parameterisations
within the resolved pomeron model.  The first two predictions are 
based on the parton distributions of the 
pomeron and sub-leading
exchange from the leading order DGLAP 
analysis of $F_2^{D}$ from H1 \cite{H1:f2d394}. 
The `ACTW fit D' parameterisation
is the best combined fit in \cite{actw} to H1 and ZEUS $F_2^{D}$
data and ZEUS diffractive dijet data.
All three sets of parton parameterisations give acceptable descriptions of $F_2^{D}$.
All of the predictions using the three parton parameterisations exceed the
data although the parameterisation with 
the flat gluon distribution (`H1 fit 2') is closest. 
The predictions shown are calculated with the factorisation and renormalisation 
scales set to 
$\mu^2 \equiv \mu_f^2 \equiv \mu_r^2 = Q^2 + p^2_{T} + 4 m_c^2$.  
Changing this scale to $ p^2_{T} + 4 m_c^2$ produces an increase of 
around $20\%$ in the predicted cross sections. Similarly, the variation of 
the charm quark mass 
by $\pm 0.1 \ {\rm GeV}$  leads to an uncertainty of $\mp 10 \%$ in the cross sections. 
Changing $\epsilon$ in the Peterson model from $0.078$ to
$0.035$ and to $0.1$ produces an uncertainty 
in the cross section predictions of $^{+15}_{-5}\%$.  The values shown in the
table are calculated with $\Lambda_{QCD}=0.20 \ {\rm GeV}$ and the number of active 
quark flavours in the first order expression for $\alpha_s$ is  
$N_f= 4$. Selecting
$\Lambda_{QCD}=0.25 \ {\rm GeV}$ and $N_f= 5$ leads to an increase of about
$10 \%$ in the cross sections.  The contribution of $\dstar$ production
from meson exchange in the predictions is less than $7\%$.

The cross section predictions from the semi-classical, SCI and GAL 
models are also shown in table \ref{table:model} and 
are in agreement with the data.  However, none of these models can
simultaneously reproduce the shapes and normalisations of the
differential
dijet cross sections\cite{H1:d2j00}. The semi-classical
model
prediction was calculated
using the same factorisation scale as for the resolved pomeron model.
The SCI and GAL model predictions use
$\mu^2 = Q^2 + 2p^2_{T} + 2 m_c^2$.  
For each of the three models, the uncertainty in the predictions due to the variation 
of the factorisation 
scale, $m_c$ and $\epsilon$ are similar to those for the resolved pomeron model.  

\begin{figure}[p]
\begin{center}
\mbox{\epsfig{figure=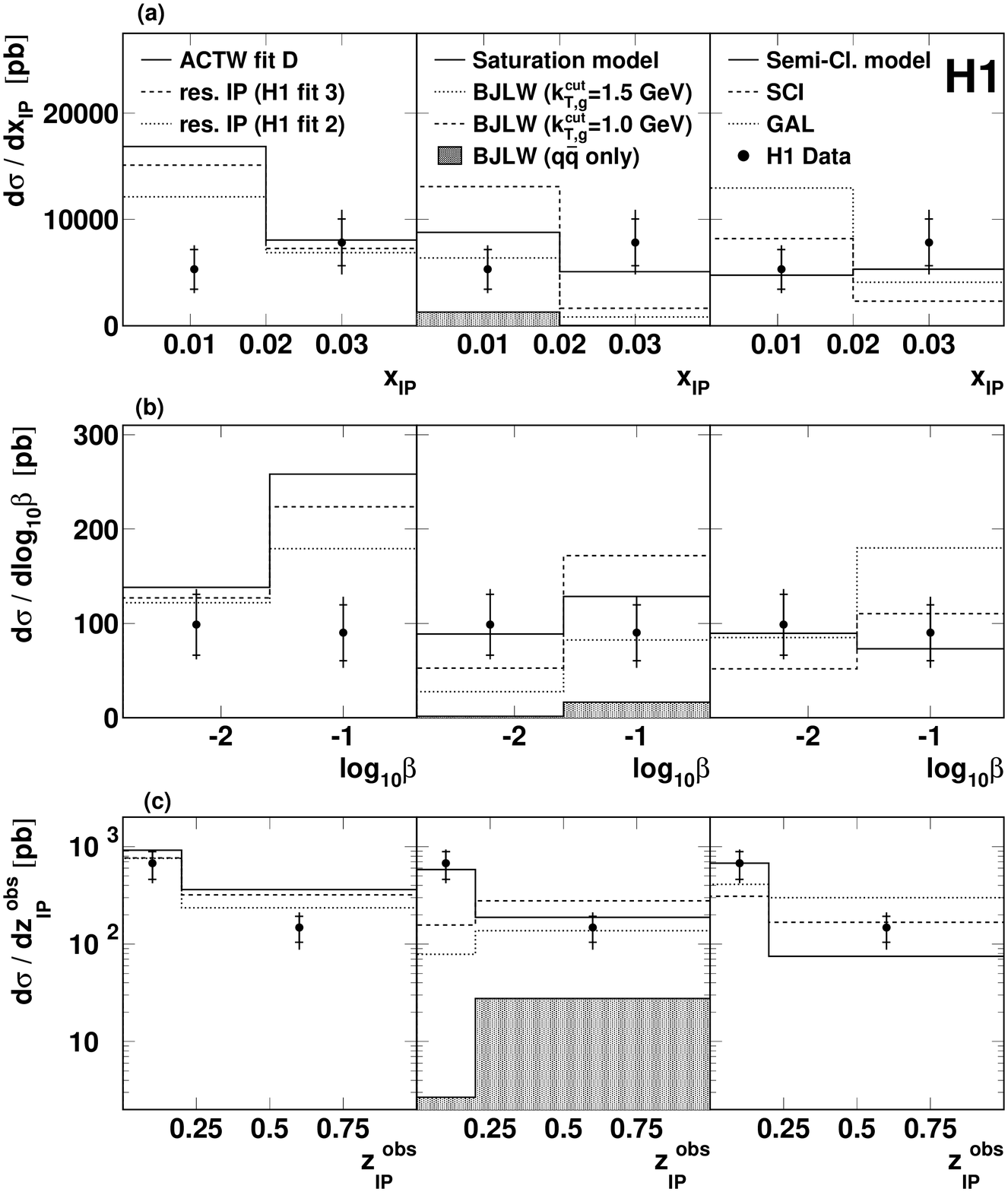,angle=0,width=1.01\textwidth}}
\end{center}
\caption{\em 
Cross sections $\sigma(ep\rightarrow e(D^{*\pm}X')Y)$ as a 
function of (a) $\xpom$; (b) $\log_{10} \beta$ and (c) $\zpomobs$.
The data are points with 
error bars (inner: statistical, outer: total).
Each distribution is plotted three times to allow comparison with 
the three groups of models described in the text.
}
\label{fig:betaxpomzpom} 
\end{figure}

\begin{figure}[p]
\begin{center}
\mbox{\epsfig{figure=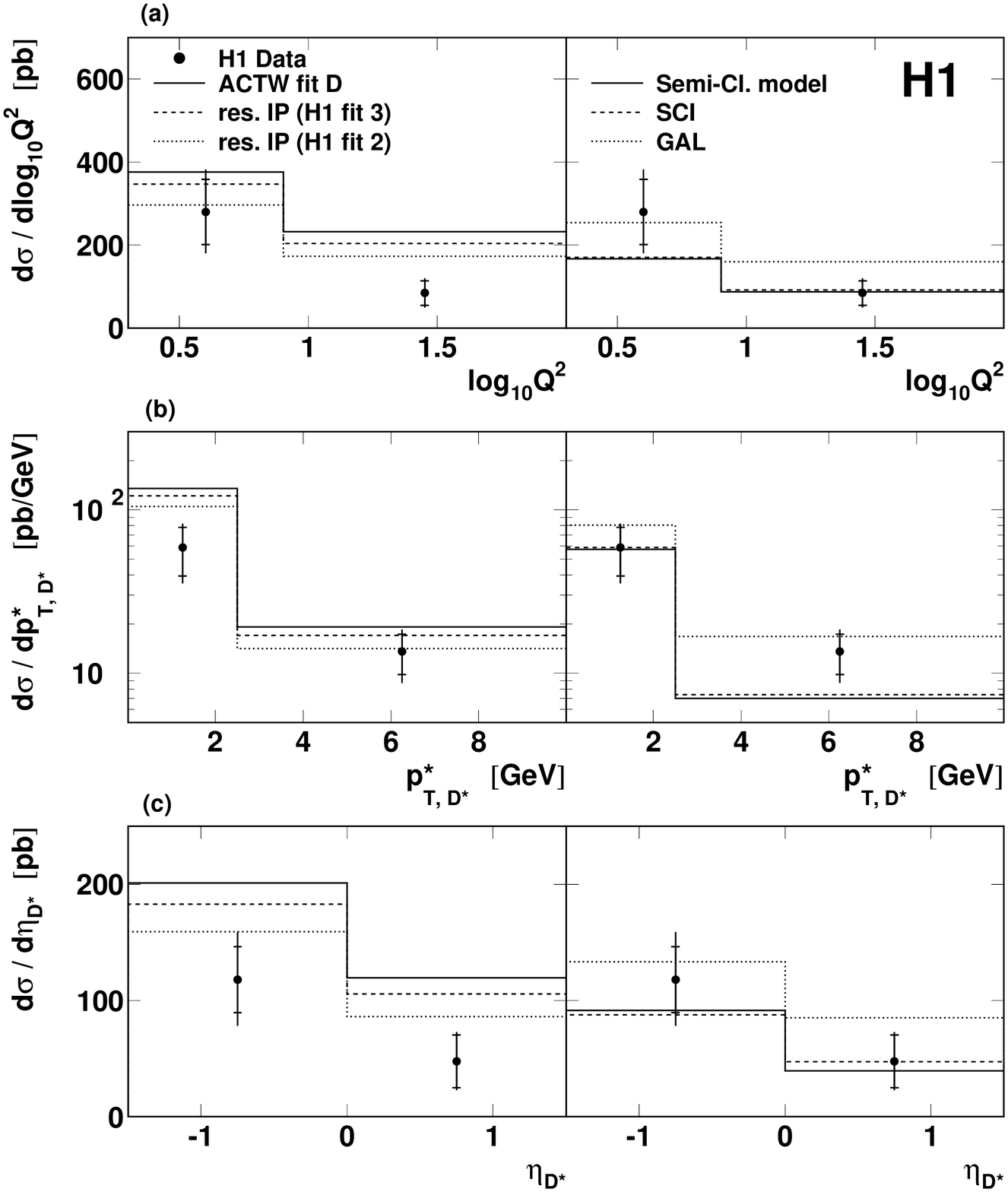,angle=0,width=1.0\textwidth}}
\end{center}
\caption{\em 
Cross sections $\sigma(ep\rightarrow e(D^{*\pm}X')Y)$ as a 
function of (a) $log_{10} Q^2$; (b) $\ptgp$
and (c) $\etalab$. 
The data are points with 
error bars (inner: statistical, outer: total).
Each distribution
is plotted twice to allow comparison with two of the
groups of models described in the text.}
\label{fig:q2ptgpeta} 
\end{figure}

Differential  cross sections are shown in
figures~\ref{fig:betaxpomzpom} and \ref{fig:q2ptgpeta}. 
They represent average values over the intervals shown in the
figures.

In figure \ref{fig:betaxpomzpom} the cross section is shown
as a function of 
$\xpom$, $\log_{10} \beta$ and $\zpomobs$ and compared to the QCD-based
models described in section \ref{section:models}.
The cross sections differential in $\xpom$ and $\log_{10} \beta$ are flat
within experimental errors.
Figure \ref{fig:betaxpomzpom} shows that about $60\%$ of charm 
production is in the region $\zpomobs < 0.2$
where the contribution of the BGF process is expected to dominate.

The discrepancy between the 
resolved pomeron model predictions and the 
data is pronounced  
in the low $\xpom$, high $\beta$ and high 
$\zpomobs$ regions which are all correlated with low values of $\mx$. Of 
the three parton parameterisations the calculations 
which use `H1 Fit 2' are shown to come closest to the data in this
region. All three parameterisations are consistent with the data in
the high $\mx$ region. 

The two gluon exchange models, directly applicable in the low $\xpom$
region ($\xpom <
0.01$) are also compared to the data in figure \ref{fig:betaxpomzpom}. 
The sensitivity of this 
measurement to the value of the transverse 
momentum cut-off $k_{T,g}^{cut}$ of the gluon in the $q\bar{q}g$ state 
within the BJLW model is 
also
studied. Calculations which use cut-off values of $1.0 \ \rm{GeV}$ and
$1.5 \ \rm {GeV}$ both give a fair description of the data in the low $\mx$
region (i.e. the low $\xpom$, high $\beta$, high $\zpomobs$ domain). 
These data also offer sensitivity to the relative 
contribution of the scattering of the 
$q\bar{q}$ fluctuation, which is shown as a shaded zone, and 
forms a sizeable component of the total two gluon exchange 
cross section.  
The saturation model
reproduces reasonably well the normalisation of the data in the
low $\xpom$ range, in which it is expected to be applicable, but also 
 provides a good
description of the data in the remaining region of phase space. 

The semi-classical model gives a good description of the distributions
shown in figure \ref{fig:betaxpomzpom}. Both the SCI and GAL models 
provide a satisfactory description of the spectra although the GAL model 
tends to overestimate the data in the low $\mx$ domain.

In figure \ref{fig:q2ptgpeta}, the $\dstar$ cross section is plotted differentially  
at 2 values of $\log_{10}Q^2$,
$p^*_{T, D^*}$ and $\eta_{D^*}$, where $p^*_{T, D^*}$ is the transverse momentum 
of the ${\dstar}$ in the $\gamma^* p$ centre of mass system. The data tend
to fall off with higher values of each of these variables. 

Since these distributions are integrated over the full $\xpom$ range of
this measurement, a 
comparison is made only with resolved pomeron
calculations and soft colour neutralisation models. Both sets of models
provide a reasonable description of the data.

\section{Conclusion}

The dynamics of open diffractive charm production in DIS have been studied 
for the first time 
at HERA.  The total $\dstar$ production cross section in the kinematic range
$2 < Q^2 < 100 ~{\rm GeV^2}$, 
$0.05 < y < 0.7$, 
$\xpom < 0.04$, 
$M_Y < 1.6 ~{\rm GeV}$, 
$|t| < 1 ~{\rm GeV^2}$, 
$\ptlab > 2 \ {\rm GeV}$ and 
$|\etalab|< 1.5$ 
has been found to be $246 \pm 54 \ {\rm (stat.)} \pm 56 \ {\rm (syst.)} \ {\rm pb}$.
In the studied region 
about $6\%$ of the total $\dstar$ cross section 
is produced diffractively.

The cross section has been measured  
as a function of $\xpom$, 
$\log_{10} \beta$, $\zpomobs$,
$\log_{10}Q^2$,
$\ptgp$ and $\etalab$. 
The data show a sizeable component of charm production in the low $\zpomobs$ 
region which is suggestive of the dominance of the contribution from the
boson--gluon--fusion process. 

A number of QCD-based models which give a good description 
of the inclusive diffractive cross section were compared with the 
measurement. 
A reasonable description of the data is provided by a model based on the
resolved pomeron picture using various assumptions for the partonic
composition of the colourless exchange. A parton parameterisation 
containing a flat gluon dependence~(`H1 Fit 2') comes closest to the data.
Predictions of two gluon
exchange processes were found to match the data in the low $\xpom$ region. 
Soft colour neutralisation models give a satisfactory description of the
data.

\section*{Acknowledgements}

We are grateful to the HERA machine group whose outstanding
efforts have made and continue to make this experiment possible. 
We thank
the engineers and technicians for their work in constructing and now
maintaining the H1 detector, our funding agencies for 
financial support, the
DESY technical staff for continual assistance
and the DESY directorate for the
hospitality which they extend to the non DESY 
members of the collaboration.


\end{document}